\documentclass[%
 prl,reprint,twocolumn,
nofootinbib,
 amsmath,amssymb,
 prl,
]{revtex4-2}

\usepackage{graphicx}
\usepackage{dcolumn}
\usepackage{bm}

\usepackage[
total={6.5in,9.75in}, top=0.9 in, left=0.9in, includefoot,
]{geometry}
\usepackage{color}

\usepackage{tikz,xcolor,hyperref}
\hypersetup{
    colorlinks,
    linkcolor={red!80!black},
    citecolor={blue!50!black},
    urlcolor={blue!80!black}
}

\definecolor{lime}{HTML}{A6CE39}
\DeclareRobustCommand{\orcidicon}{
	\begin{tikzpicture}
	\draw[lime, fill=lime] (0,0)
	circle [radius=0.16]
	node[white] {{\fontfamily{qag}\selectfont \tiny ID}};
	\draw[white, fill=white] (-0.0625,0.095)
	circle [radius=0.007];
	\end{tikzpicture}
	\hspace{-2mm}
}

\foreach \x in {A, ..., Z}{\expandafter\xdef\csname orcid\x\endcsname{\noexpand\href{https://orcid.org/\csname orcidauthor\x\endcsname}
			{\noexpand\orcidicon}}
}

\begin{document}

\preprint{APS/123-QED}

\title{Elastic frontogenesis sets the spectra of smooth elastic turbulence}

\author{Marco Edoardo Rosti\orcidA{}}
\email{marco.rosti@oist.jp}
\affiliation{Complex Fluids and Flows Unit, Okinawa Institute of Science and Technology Graduate University, Okinawa 904-0495, Japan}
\author{Andrea Mazzino\orcidB{}}
\email{andrea.mazzino@unige.it}
\affiliation{Department of Civil, Chemical and Environmental Engineering, University of Genova, Via Montallegro 1, 16145 Genova, Italy \\
            Istituto Nazionale di Fisica Nucleare, Sezione di Genova, Via Dodecaneso 33, 16146 Genova, Italy
            }

\begin{abstract}
Most turbulent flows are sustained by inertia, which transfers energy across scales and produces characteristic power-law spectra. Yet dilute polymer solutions also exhibit turbulence when inertia is negligible: in this state, known as elastic turbulence, the velocity field remains spatially smooth while both the velocity and the polymer conformation exhibit robust power-law spectra. How a smooth flow, lacking a hierarchy of eddies, generates such multiscale structure has remained unexplained. Here we show that the missing mechanism is elastic frontogenesis, whereby the flow spontaneously organizes polymer stress into thin sheets that are stretched within their plane and compressed across their thickness, causing their amplitude to grow as they become thinner. Because the fronts are sheet-like structures, this amplification–thinning law uniquely determines the polymer conformation spectrum, while the Stokes response transfers the same geometric scaling to the velocity field, reconciling broadband spectra with spatial smoothness. Being geometric rather than model-specific, elastic frontogenesis provides an organizing principle for viscoelastic flows and the viscoelastic analogue of vortex stretching in inertial turbulence.
\end{abstract}

\maketitle


Turbulent flows are characterized by universal power-law spectra. In classical, inertia-dominated turbulence, this multiscale structure is organized by the energy cascade: nonlinear advection transfers energy across a hierarchy of eddies, imprinting the celebrated Kolmogorov scaling on the velocity field \cite{kolmogorov_1941a, frisch_1995k}. Whether and how universal spectra can emerge in a flow with no cascade at all is a fundamental question in non-equilibrium physics.

Dilute polymer solutions provide a striking arena for this question. When elastic stresses dominate over inertia, these fluids sustain a chaotic flow state known as elastic turbulence (ET) \cite{groisman_steinberg_2000a, benzi_ching_2018f, steinberg_2021p}. Originally discovered in flows with curved streamlines \cite{shaqfeh_1996a, datta_ardekani_arratia_beris_bischofberger_mckinley_eggers_lopez-aguilar_fielding_frishman_others_2022a}, ET has since been observed in homogeneous, wall-bounded, multiphase and porous flows\cite{burghelea_segre_steinberg_2007a, berti_bistagnino_boffetta_celani_musacchio_2008a, yerasi_picardo_gupta_vincenzi_2024a, singh_perlekar_mitra_rosti_2024a, zhang_li_fan_su_xi_sun_2025a, bonn_ingremeau_amarouchene_kellay_2011a, qin_arratia_2017s, lellep_linkmann_morozov_2024a, foggi-rota_amor_le-clainche_rosti_2024a, yamani_keshavarz_raj_zaki_mckinley_bischofberger_2021p, yamani_raj_zaki_mckinley_bischofberger_2023a, soligo_rosti_2023a, walkama_waisbord_guasto_2020z, browne_datta_2021a, haward_hopkins_shen_2021a, blois_haward_shen_2023a, lopez-de-la-cruz_haward_shen_2025a, sun_xue_aime_kim_tang_mckinley_stone_weitz_2023a, ravisankar_zenit_2025a}, and even within the small scales of inertial polymeric turbulence \cite{garg_rosti_2025a}, demonstrating that it is a remarkably robust state of viscoelastic fluids. Across these settings, experimental and numerical studies have converged on an apparently universal decay of the spatial kinetic-energy spectrum, $E_f\left(k\right) \sim k^{-4}$ \cite{berti_bistagnino_boffetta_celani_musacchio_2008a, berti_boffetta_2010a, ray_vincenzi_2016a, lellep_linkmann_morozov_2024a, singh_perlekar_mitra_rosti_2024a, beneitez_page_dubief_kerswell_2024a, singh_rosti_2025a, serafini_2025a, garg_rosti_2025a}, while simulations consistently find that the spectrum of the polymer conformation follows $E_p\left(k\right) \sim k^{-3/2}$ \cite{singh_perlekar_mitra_rosti_2024a, beneitez_page_dubief_kerswell_2024a, singh_rosti_2025a, serafini_2025a}.

The coexistence of these two laws poses a puzzle. Elastic turbulence develops when inertia is negligible: the velocity field remains spatially smooth at leading order, with intermittency confined to subleading corrections \cite{singh_perlekar_mitra_rosti_2024a}, and there is no advective cascade to organize the dynamics scale by scale. The seminal theory of Fouxon and Lebedev \cite{fouxon_lebedev_2003k} established that this is not, by itself, an obstruction: a smooth, elastically driven flow can sustain power-law spectra without any flux of a conserved quantity, with exponents determined implicitly by the Lagrangian statistics of the large-scale flow. What that theory leaves open is the \textit{selection problem}: what real-space object replaces eddies as the carrier of scale-by-scale structure, and what mechanism fixes the robust exponents observed across ET flows?

A natural first guess is that the spectra are controlled by the most stretched polymers. This intuition, however, conflates two distinct notions: amplitude and scale. A peak of polymer stretch identifies where elastic deformation accumulates, whereas a spectrum describes how that deformation is distributed across scales. The relevant structures are therefore not the maxima of polymer stretching, but the thin transition layers across which the polymer stress changes sharply in space, and it is their geometry, rather than the magnitude of the stretching, that determines how elastic stresses populate different scales.

\begin{figure}[t]
	\includegraphics[width=0.49\textwidth]{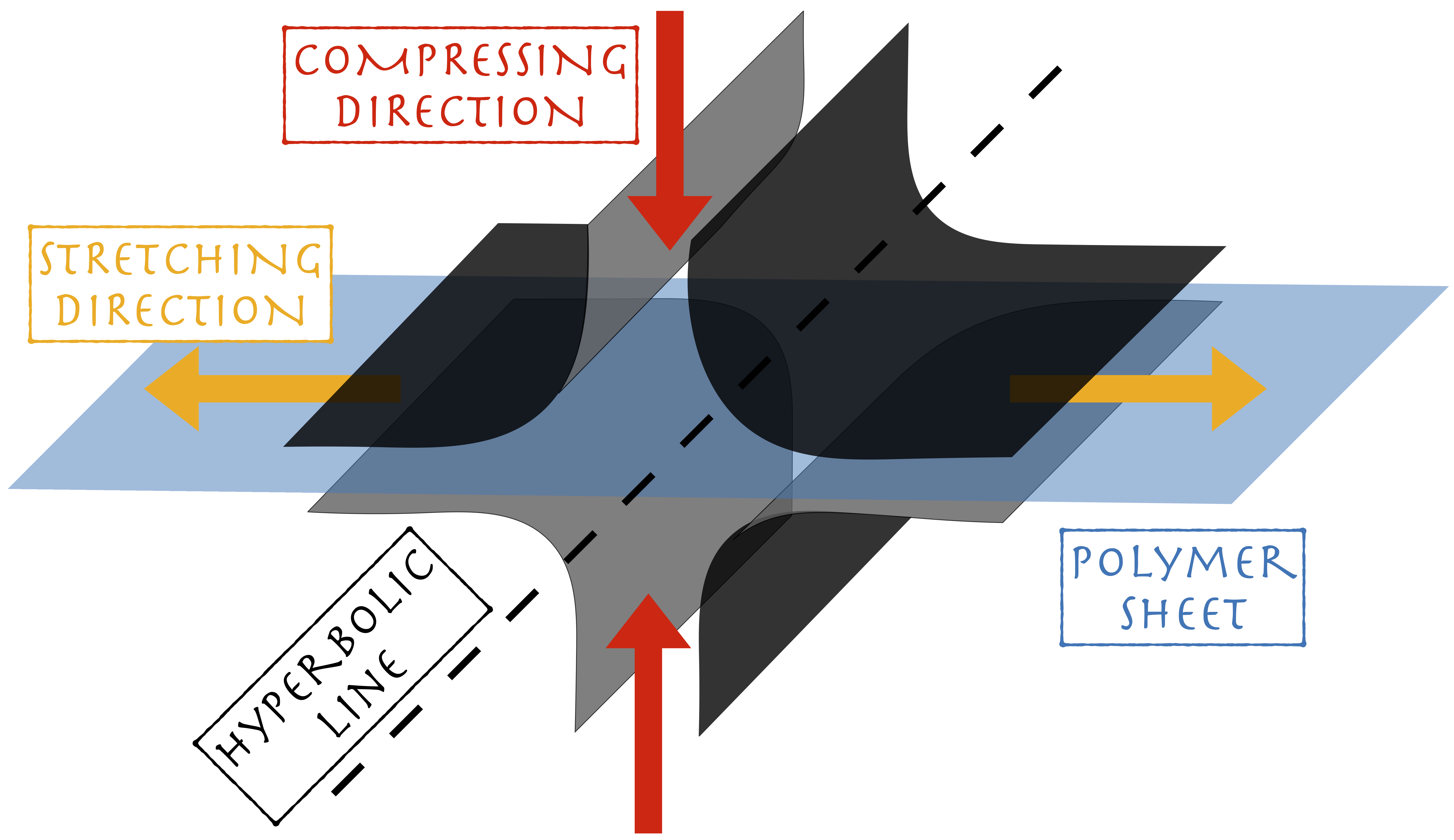}
	\caption{\textbf{Elastic frontogenesis: a sketch of the mechanism.} In the regions of steepest polymer gradients, the local strain field is of near-plane-strain type: denoting by $s_1\ge s_2\ge s_3$ the principal strain rates (the eigenvalues of the symmetric part of the velocity-gradient tensor), the flow is stretched in one direction ($s_1=s>0$), compressed in another ($s_3=-s<0$), and nearly neutral in the third ($s_2\approx 0$). 
	Material is stretched tangentially while being squeezed across its thickness, so that the polymer stress organizes into a thin sheet lying in the plane spanned by the stretching and neutral directions, with its normal along the compressive direction; the sheet's amplitude grows while its thickness shrinks.}
	\label{fig:fig3}
\end{figure}

Here we show, combining direct numerical simulations with a geometric theory, that these layers are generated by a mechanism we term \textit{elastic frontogenesis}. In the regions where polymer gradients are steepest, the local flow takes a near-plane-strain form: one principal direction stretches, one compresses, and one remains nearly neutral (Fig.~\ref{fig:fig3}). Such a flow naturally manufactures sheets: material elements are stretched tangentially while being squeezed along the compressive axis, so that the polymer stress organizes into thin sheet-like fronts whose normals align with the compressive direction, and whose amplitude grows while their thickness shrinks.

We show that the coupled growth of sheet amplitude and thinning converts a simple geometrical law into the universal spectra of elastic turbulence. The codimension-one geometry of the sheets directly yields the polymer-conformation spectrum, while the Stokes balance between polymer forcing and viscous stresses fixes the velocity spectrum. Elastic frontogenesis therefore resolves the long-standing selection problem by identifying the physical structures that organize multiscale dynamics in elastic turbulence. More broadly, these results suggest that elastic frontogenesis represents the fundamental analogue of vortex stretching in inertial turbulence, providing the organizing principle that links coherent structures to universal spectral scaling in elastic turbulence.


We now establish this picture, starting from the equations governing the coupled dynamics of the velocity and polymer fields,
\begin{equation}
\begin{split}
\rho_f \left( \partial_t \boldsymbol{u} +  \left( \boldsymbol{u} \cdot \boldsymbol{\nabla} \right) \boldsymbol{u} \right) =-\boldsymbol{\nabla} p + \mu_f \nabla^2 {\boldsymbol{u}} +\\
(\mu_p/\tau_p) \boldsymbol{\nabla} \cdot \left( f_1 \boldsymbol{C}- f_2 \textbf{I} \right) + \boldsymbol{F_{ext}},
\end{split}
\end{equation}
\begin{equation}
\begin{split}
\partial_t \boldsymbol{C} + \boldsymbol{u} \cdot \boldsymbol{\nabla} \boldsymbol{C} =\\
\boldsymbol{C} \cdot \boldsymbol{\nabla}  \boldsymbol{u} + \boldsymbol{\nabla} \boldsymbol{u}^{T} \cdot \boldsymbol{C} - \frac{1}{\tau_p} \left( f_1 \boldsymbol{C} - f_2 \textbf{I} \right).
\end{split}
\end{equation}
Here $\boldsymbol{C}$ is the polymer configuration field, a positive-definite second-order tensor measuring the local size and orientation of the polymer coils: it is close to the identity when polymers are relaxed, and develops at least one large eigenvalue when they are stretched. In the momentum equation, $\rho_f$ and $\mu_f$ are the fluid density and viscosity, $p$ the pressure, $\mu_p$ the polymeric viscosity, $\tau_p$ the polymer relaxation time, $\textbf{I}$ the identity tensor, and $\boldsymbol{F_{ext}}$ the external forcing sustaining the flow (see End Matter), while $f_1$ and $f_2$ are terms distinguishing various polymeric models. For example, $f_1 = 1$ and $f_2=1$ for the Oldroyd-B model, and $f_1 = \frac{L_{m}^2}{L^2_{m}-\mathrm{tr}(\boldsymbol{C})}$ and $f_2=\frac{L_{m}^2}{L^2_{m}-3}$ for the FENE-P model, with $L_m$ the maximum polymer extensibility \cite{oldroyd_1950a, peterlin_1966a, bird_armstrong_hassager_1987a, bird_curtiss_armstrong_hassager_1987a, larson_1988a}.

\begin{figure}[th]
	\includegraphics[width=0.49\textwidth]{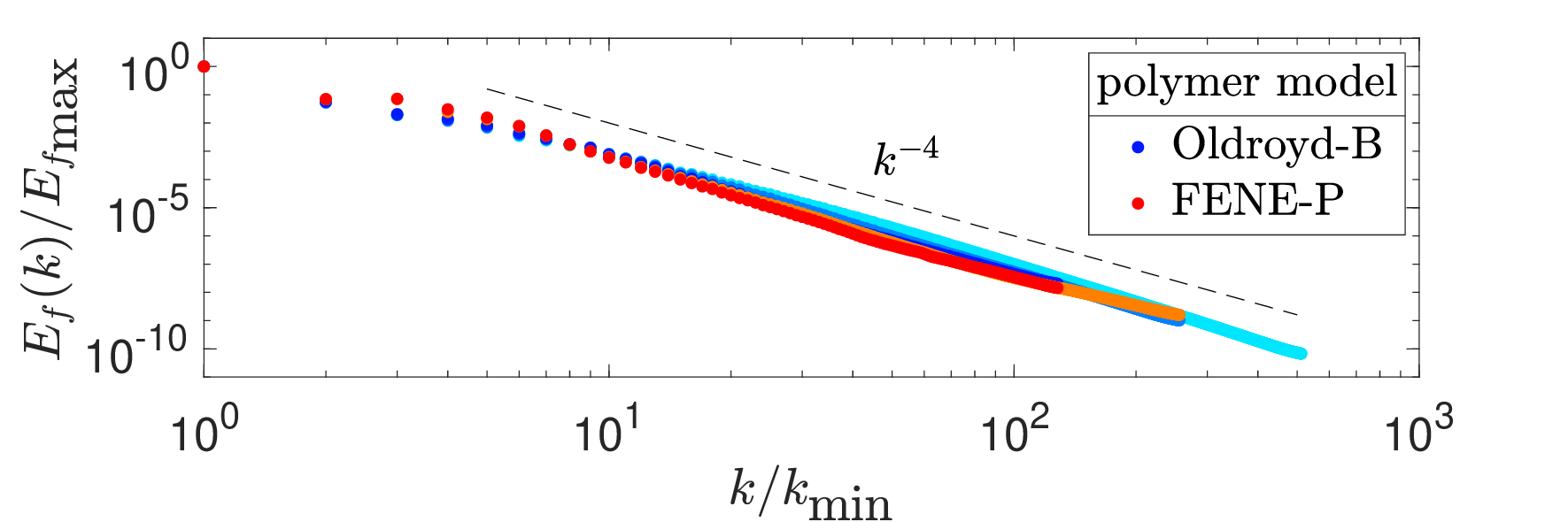}
	\includegraphics[width=0.49\textwidth]{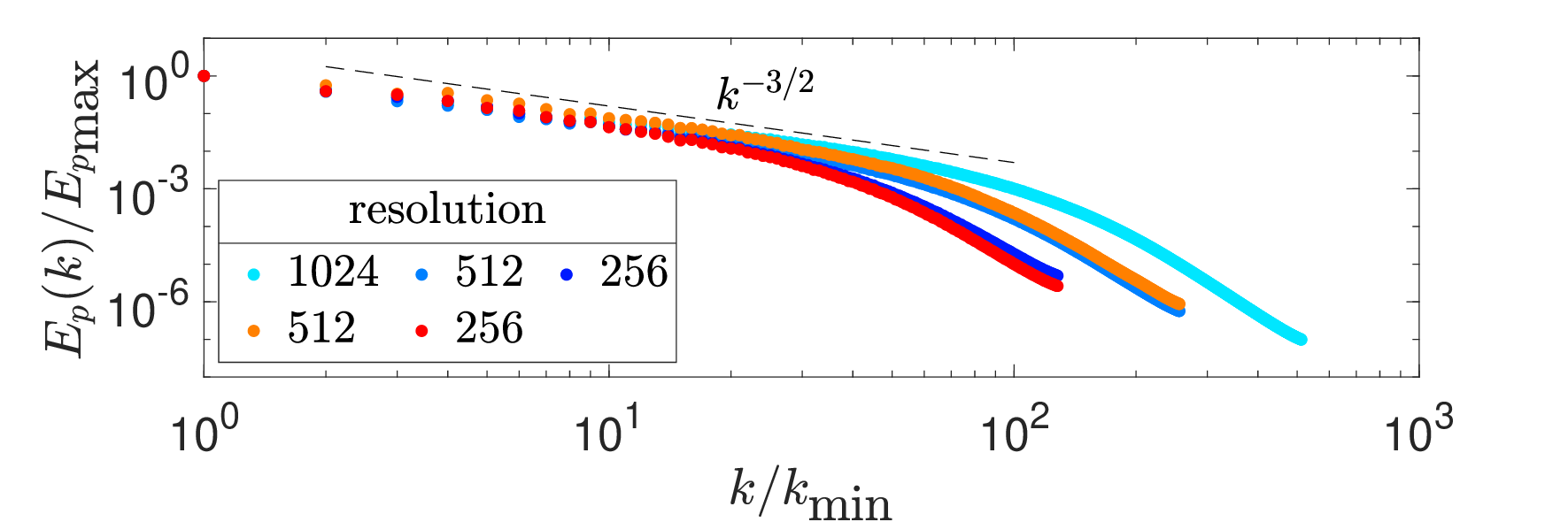}	
	\caption{\textbf{The two spectral laws of smooth elastic turbulence.} The energy spectrum of the (top) fluid $E_f \left( k \right)$ and (bottom) polymer $E_p \left( k \right)$ versus the wavenumber $k$, obtained from our simulations. The simulations are performed with the Oldroyd-B (blue) and FENE-P (red) models, in a cubic triperiodic domain. See End Matter for additional information on the numerical simulations, and the Supplementary Material for the role of inertia \cite{supplementary}.}
	\label{fig:fig1}
\end{figure}

The two spectra to be explained are built from these fields. For the velocity fluctuations, the energy spectrum is defined as
\begin{equation}
  E_f \left( k \right) \equiv \int_{\lvert \mathbf{k}\rvert=k} \langle \widehat{u}_{i} \left( \mathbf{k},t \right) \widehat{u}_{i}^* \left( \mathbf{k},t \right) \rangle dS_k, \label{eq:Ek}
\end{equation}
where $dS_k$ is the surface measure on the sphere $\lvert \mathbf{k}\rvert=k$, and $\widehat{u}_i \left( \mathbf{k},t \right)$ is the Fourier transform of the velocity field $u_i \left( \mathbf{x},t \right)$. Analogously, the multiscale content of the polymer field is characterized by the conformation spectrum
\begin{equation}
  E_p \left( k \right) \equiv \int_{\lvert \mathbf{k}\rvert=k} \langle \widehat{B}_{ij} \left( \mathbf{k},t \right) \widehat{B}_{ij}^* \left( \mathbf{k},t \right) \rangle dS_k, \label{eq:Ep}
\end{equation}
where the matrix $\boldsymbol{B}$ is the symmetric positive-definite square root of $\boldsymbol{C}$, defined by $C_{ij} = B_{il}B_{lj}$ \cite{balci_thomases_renardy_doering_2011l, nguyen_delache_simoens_bos_el-hajem_2016t}. While $E_p$ cannot at present be measured experimentally, owing to the inability to visualize the conformation-tensor field, it is readily accessible in simulations. Figure~\ref{fig:fig1} shows both spectra as obtained from our direct numerical simulations of ET in a triperiodic box sustained by large-scale forcing, for the Oldroyd-B and FENE-P models, together with the two power laws $E_f \sim k^{-4}$ and $E_p \sim k^{-3/2}$ (see End Matter and the Supplementary Material \cite{supplementary}). \textit{Where, in real space, is this scale-by-scale structure assembled?} Note that, these spatial exponents should not be conflated with the temporal ones measured at fixed probes in space, which scatter widely across experiments, as recently reviewed in \cite{foggi-rota_singh_chiarini_amor_soligo_mitra_rosti_2026a}. 

\paragraph{Amplitude is not scale}
We first put the distinction between amplitude and scale on quantitative grounds, by following the polymers along their trajectories. Along a trajectory \cite{kumar_guasto_ardekani_2023a}, the transport equation for $\boldsymbol{C}$ becomes
\begin{equation}
D_t \boldsymbol{C} = \boldsymbol{C} \cdot \boldsymbol{\nabla}  \boldsymbol{u} + \boldsymbol{\nabla} \boldsymbol{u}^{T} \cdot \boldsymbol{C} - \frac{1}{\tau_p} \left( f_1 \boldsymbol{C} - f_2 \textbf{I} \right),
\end{equation}
where $D_t$ is the material derivative. Taking the trace of this equation gives
\begin{equation} \label{eq:DTRC}
D_t \mathrm{tr}(\boldsymbol{C}) = 2 \boldsymbol{S} : \boldsymbol{C} - \frac{1}{\tau_p} \left( f_1 \mathrm{tr}(\boldsymbol{C}) - f_2 3 \right),
\end{equation}
where $\boldsymbol{S} : \boldsymbol{C} \equiv S_{ij}C_{ij}$, and $\boldsymbol{S}$ is the symmetric part of the gradient of the velocity field tensor, i.e., $\boldsymbol{S} = \left( \boldsymbol{\nabla} \boldsymbol{u} + \boldsymbol{\nabla} \boldsymbol{u}^{T} \right)/2$. We can thus define a scalar measure of polymer stretching as $R(\bm x,t) \equiv \mathrm{tr}(\boldsymbol{C}) -3$: $R=0$ is the isotropic equilibrium state, while $R>0$ measures excess stretching. We now divide Eq.~\eqref{eq:DTRC} by $R(\bm x,t)$ and obtain
\begin{equation}
g \equiv D_t \mathrm{ln} R = 2 q - \frac{1}{\tau_p} \frac{f_1 \mathrm{tr}(\boldsymbol{C}) - f_2 3}{\mathrm{tr}(\boldsymbol{C}) -3},
\end{equation}
where we can interpret $g$ as the logarithmic growth rate of the polymer amplitude, and $q \equiv \boldsymbol{S} : \boldsymbol{C}/R$ as an effective stretching rate. Positive $g$ means that the polymer amplitude is growing, whereas negative $g$ means that it is decaying.
\begin{figure}[t]
	\includegraphics[width=0.49\textwidth]{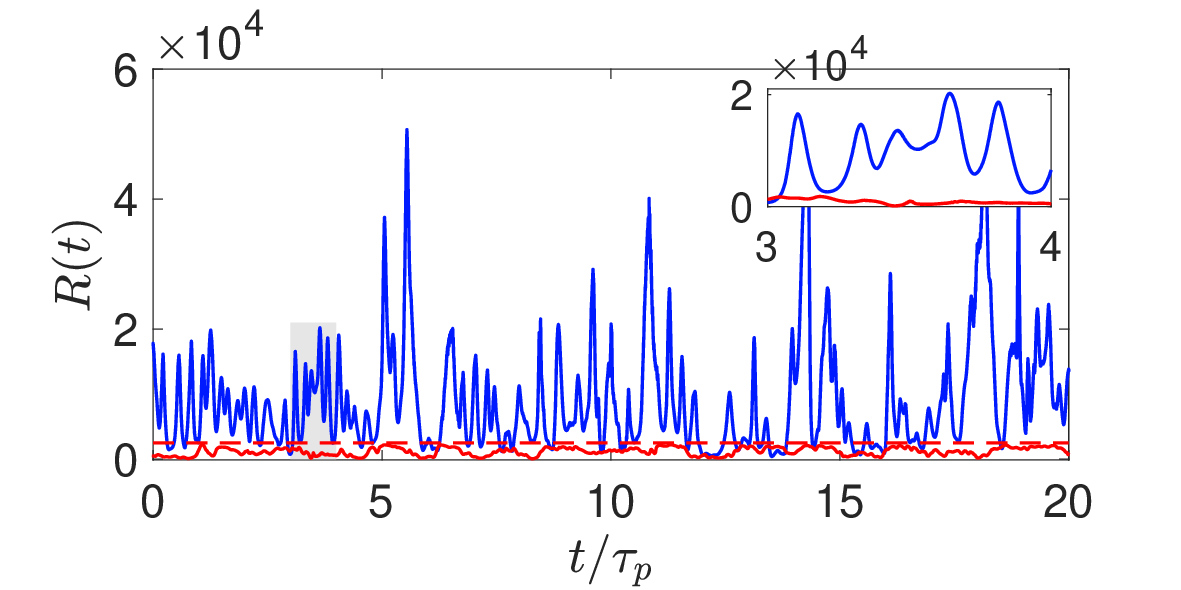}
	\includegraphics[width=0.49\textwidth]{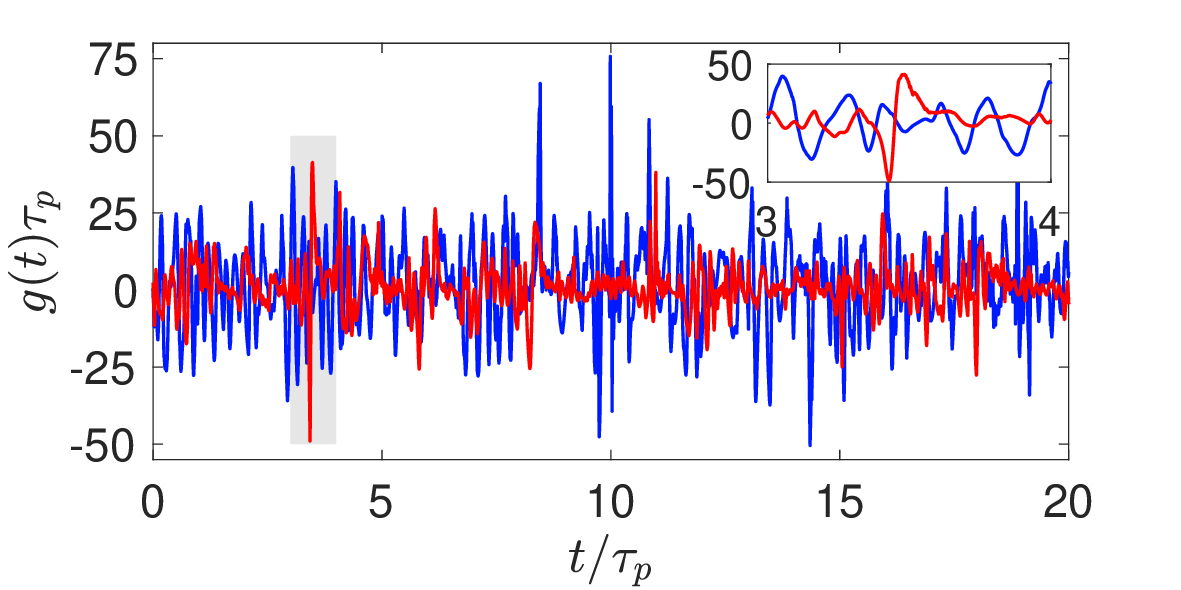}
	\caption{\textbf{Verification of elastic frontogenesis.}
	\textbf{Peaks of polymer stretching are turning points, not carriers of scale.} Time histories of the (top) polymer stretching $R(t)$ and (bottom) its logarithmic growth rate $g(t)$ along a tracer, with the zeros of $g$ marking turning points of $R$. The blue and red lines are used for the results from the Oldroyd-B and FENE-P models, respectively, with the red dashed horizontal line in the left panel marking the maximum allowed stretching of the polymers $L_m^2-3$ for the FENE-P model. The insets of the two figures show enlarged views of the gray rectangles, spanning a period of time equal to one relaxation time in the range $t/\tau_p\in \left[ 3-4 \right]$.}
	\label{fig:fig2bis}
\end{figure}
\begin{figure*}[t]
	\includegraphics[width=0.32\textwidth]{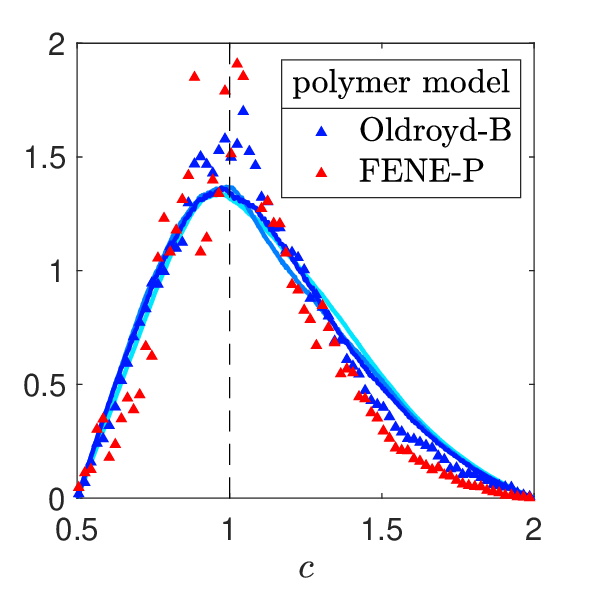}
	\includegraphics[width=0.32\textwidth]{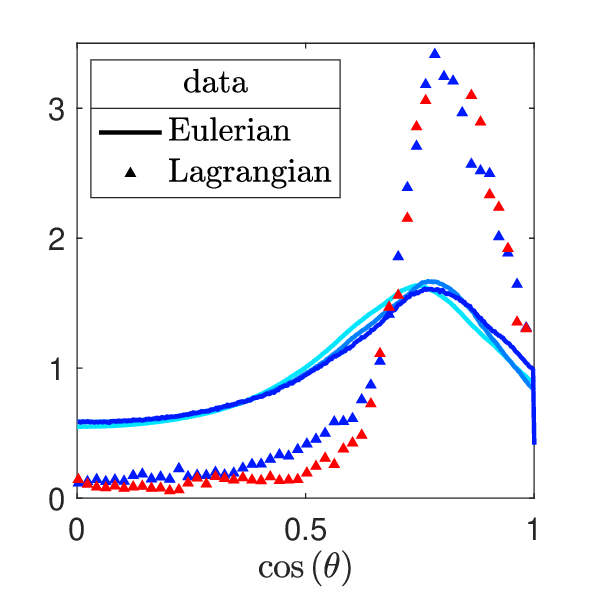}
	\includegraphics[width=0.32\textwidth]{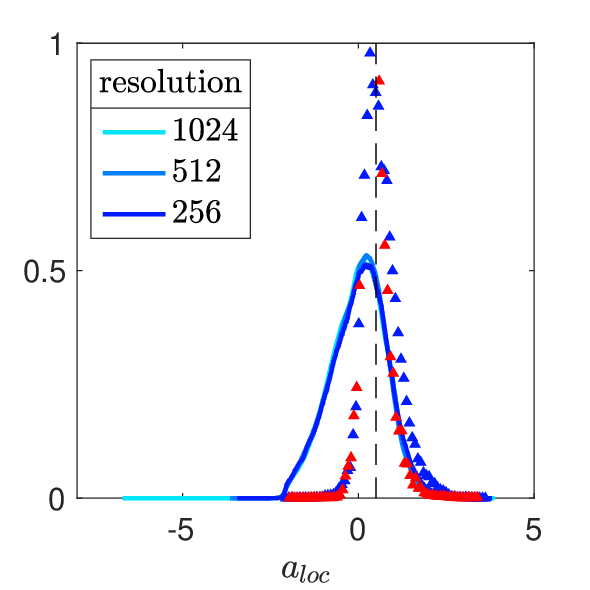}
	\caption{\textbf{Verification of elastic frontogenesis.}
	\textbf{The three signatures of elastic frontogenesis.} Probability distribution functions (pdf) of the local diagnostics. (left) The ratio $c=-s_3/s_1$ indicates the strain geometry, with the data supporting near plane strain geometry ($c \simeq 1$, vertical dashed line). (middle) The alignment $\cos \left( \theta \right) = |\hat{\bm n}\cdot\bm e_3|$ of the front normal with the compressive direction. The distribution is skewed towards $1$, confirming the alignment of the two vectors. (right) The growth to thinning ratio $a_{\rm loc}$, with a clear peak at $\simeq 1/2$ (vertical dashed line). The blue and red colors are used to distinguish the results from the Oldroyd-B and FENE-P models, respectively. The lines are unconditioned Eulerian data, with color tone indicating different grid resolutions $N$. The symbols indicate Lagrangian data conditioned on $\mathcal S_p$, with threshold $\Theta_p$ chosen as the quantile 0.99 ($\bigtriangleup$). Statistics are accumulated over 10 independent Eulerian snapshots and $10^5$ Lagrangian tracers, see End Matter for more details. The same quantities evaluated on $\mathcal S_f$ are reported in the Supplementary Material \cite{supplementary}.}
	\label{fig:fig2}
\end{figure*}
The tracer histories in Fig.~\ref{fig:fig2bis} show that $g$ changes sign at least twice during a macro event of large $R$. At a local maximum of $R$, $g=0$ by definition (and consequently $2 q\tau_p = \left( f_1 \mathrm{tr}(\boldsymbol{C}) -f_2 3 \right)/\left( \mathrm{tr}(\boldsymbol{C}) -3 \right)$). This confirms that peaks of $R$ are turning points of the amplitude of polymer stretching, but they are not, by themselves, the relevant part to look into to understand the mechanism that creates the polymer conformation spectrum. Note also that, while the dynamic of $R$ is markedly different among the two models considered, with the results from the FENE-P model being bounded by $L_m^2-3$, this is not the case for that of $g$. The relevant structures must thus be sought where the polymer field has steep spatial gradients.

\paragraph{Polymer-stress sheets and the amplitude--thickness law}
The polymer conformation spectrum is built from the tensor $\boldsymbol{B}$ introduced in Eq.~\eqref{eq:Ep}; it is therefore most directly linked to regions where $\boldsymbol{B}$ changes sharply in space, and a natural diagnostic is $|\boldsymbol{\nabla} \boldsymbol{B}|^2$. The velocity spectrum is linked to polymer gradients through the divergence of the polymer stress $\boldsymbol{\nabla} \cdot f_1 \textbf{C} $, and thus a diagnostic based on $|\boldsymbol{\nabla} \cdot f_1 \boldsymbol{C} |^2$ identifies polymer-stress structures that directly force the velocity at small scales. We therefore use these diagnostics for the polymer and velocity sides of the problem, and together they identify sharp structures in the field from which the spectra are built, with both pointing to \textit{polymer-stress sheets}. To prove this, we define two strong gradient sets, the first tied to the polymer conformation spectrum $\mathcal S_p=\{|\boldsymbol{\nabla} \boldsymbol{B}|^2>\Theta_p \}$, and the second tied to the polymer forcing of the velocity $\mathcal S_f=\{ |\boldsymbol{\nabla} \cdot \left( f_1 \boldsymbol{C} \right) |^2>\Theta_f \}$, with the thresholds $\Theta_p$ and $\Theta_f$ selecting strong gradients.


Let $s_1\ge s_2\ge s_3$ be the eigenvalues of $\boldsymbol{S}$, and let $\bm e_3$ be the eigenvector associated with the most compressive eigenvalue $s_3$. Since the flow is incompressible, we also have that $s_1+s_2+s_3=0$. Define
\begin{equation}
c \equiv - \frac{s_3}{s_1}=1+\frac{s_2}{s_1}.
\end{equation}
In the strong gradient set $\mathcal S_p$ (as well as in $\mathcal S_f$, see the Supplementary Material \cite{supplementary}), the data in Fig.~\ref{fig:fig2}-left are consistent with $\boxed{c\simeq 1}$, which means $(s_1,s_2,s_3)\simeq (s,0,-s)$. The local flow has thus one stretching direction, one compressing direction, and one almost neutral direction, which is a near plane strain hyperbolic geometry. The picture is intuitive: there is a local hyperbolic line along which the deformation is weak, while in the cross plane normal to it, material is stretched in one direction and compressed in the other, as sketched in Fig.~\ref{fig:fig3}. A cartoon is provided by counter rotating rollers: in the cross section there is a saddle, while along the roller axes the strain is nearly neutral. Such a flow creates a sheet, and a sheet has a normal direction. For a front detected through $\boldsymbol{B}$, an intrinsic normal $\hat{\bm n}$ can be obtained from the principal eigenvector of $M_{ij}=\partial_i B_{ab}\,\partial_j B_{ab}$, which gives the direction of strongest variation of $\boldsymbol{B}$.  In the strong gradient sets, DNS data in Fig.~\ref{fig:fig2}-middle shows $\cos \left( \theta \right) \equiv |\hat{\bm n}\cdot \bm e_3|$, with values biased toward one, i.e., $\boxed{\cos \left( \theta \right) \gg 1/2}$.  Since $1/2$ is the isotropic reference value for $\cos \left( \theta \right)$, this means that the front normal aligns with the compressive strain direction.  Mechanically, this is exactly how a material sheet is made: compression squeezes the sheet thickness, while extension stretches it tangentially.

We now ask how a compressed front produces a spectral exponent.  Follow a material trajectory $\bm X_f(t)$ embedded in a front.  Define $R_f(t)=R(\bm X_f(t),t)$ and, consequently, $\boldsymbol{C}_f=\boldsymbol{C} (\bm X_f(t),t)$ and $q_f=q(\bm X_f(t),t)$. The amplitude equation is
\begin{equation} \label{eq:eqA}
\frac{d}{d t}\ln R_f(t)=g_f(t) =2q_f-\frac{1}{\tau_p} \frac{f_1 \textrm{tr} \boldsymbol{C}_f-f_2 3}{\textrm{tr} \boldsymbol{C}_f-3}.
\end{equation}
Let $\ell_f(t)$ be the thickness of the front, measured along its normal direction.  Since the normal aligns with $\bm e_3$, local linear kinematics gives
\begin{equation} \label{eq:eqB}
\frac{d}{d t}\ln \ell_f^{-1}(t) = -s_3(\bm X_f(t),t).
\end{equation}
Equation~\eqref{eq:eqA} tells how fast the polymer amplitude grows, while Eq.~\eqref{eq:eqB} tells how fast the front becomes thinner. Their dimensionless ratio is
\begin{equation}
a_{\rm loc}(t)
\equiv
\frac{\displaystyle \frac{d}{d t}\ln R_f(t)}
{\displaystyle \frac{d}{d t}\ln \ell_f^{-1}(t)}
=
\frac{g_f (t) }{-s_3(\bm X_f(t),t)} .
\end{equation}
This number answers a simple mechanical question: when compression makes the front thinner, how much does the polymer amplitude grow? The DNS data in Fig.~\ref{fig:fig2}-right show that the values of $a_{\rm loc}$ measured on 
strong-gradient events are concentrated around a typical value, which we denote by $\bar{a}$. This means that, for the front-forming events that dominate the statistics, the two logarithmic rates are approximately proportional:
\begin{equation}
\frac{d}{d t}\ln R_f \simeq \bar{a} \frac{d}{d t}\ln \ell_f^{-1}.
\end{equation}
Now we can eliminate the time variable, not because it is irrelevant, but because the increase of amplitude can be expressed directly in terms of the decrease of thickness. If two fronts have different histories but reach the same thickness $\ell_f$, the mechanism assigns them a typical amplitude through the same amplitude-thickness relation. Thus,
\begin{equation}
\frac{d \ln R_f}{d \ln \ell_f^{-1}}\simeq \bar{a},
\end{equation}
and integrating gives $\boxed{R_f(\ell_f)\sim \ell_f^{-\bar{a}} }$, which means that, the thinner the front is, the stronger is the polymer amplitude inside it.
DNS indicates that $\boxed{a_{\rm loc}\simeq \frac12 }$ on the strong-gradient events, see Fig.~\ref{fig:fig2}-right. Thus, $R_f(\ell)\sim \ell^{-1/2}$, with the amplitude growing, but only moderately: it increases at one half of the logarithmic thinning rate. See the Supplementary Material \cite{supplementary} for an instantaneous visualisation of the local diagnostics conditioned on $\mathcal S_p$, also visually confirming the presence of polymer sheets.

\paragraph{From sheets to spectra}
We can now turn this front law into a spectrum.  In strongly stretched regions the squared amplitude of $\boldsymbol{B}=\boldsymbol{C}^{1/2}$ is proportional, up to order one geometrical factors, to $R$.  Across the normal direction of a front, the amplitude-thickness law gives $(\delta B)^2\sim R_f(\ell)\sim \ell^{-\bar{a} }$. Here the separation $r$ used in the difference $\delta$ should not be interpreted as the lateral size of a sheet, rather the scale $r$ is the distance over which the field is probed.  If an increment of length $r$ crosses a sheet, it samples the normal contrast of $\boldsymbol{B}$ across that front. The sheet contribution to the second-order structure function can therefore be written as $S_2^B(r) \equiv \langle |B(\bm x+\bm r)-B(\bm x)|^2 \rangle \sim \mathbb P(\text{crossing at scale }r)\, \left\langle |\delta_r B|^2\,\middle|\,\text{crossing of a sheet}\right\rangle$. For a codimension-one sheet, the crossing probability is proportional to $r$, and the thinning law gives the conditional squared contrast as $\left\langle |\delta_r B|^2\,\middle|\,\text{crossing of a sheet}\right\rangle \sim r^{-\bar{a} }$. Hence, $S_2^B(r)\sim r\,r^{-\bar{a} }=r^{1-\bar{a} }$. For a shell integrated spectrum $E_p(k)\sim k^{-\chi}$ (with $1<\chi<3$), the usual relation between spectrum and second order increments is $ S_2^B(r)\sim r^{\chi-1}$. Hence, $\chi-1=1-\bar{a} $, from which we obtain $\boxed{\chi=2-\bar{a}}$. With $\bar{a} =1/2$, the exponent of the polymer conformation spectrum results in $\chi=\frac32$ and therefore $E_p \left( k \right) \sim k^{-3/2}$. Following \cite{singh_perlekar_mitra_rosti_2024a} (and generalised to different models in the End matter), we finally obtain $\boxed{E_f(k)\sim E_p^2(k)/k}$.

\paragraph{Discussion and conclusions}
In this work we provide a theoretical derivation for the spectral decay rates of the fluid $E_f$ and polymer conformation $E_p$, based on elastic frontogenesis. The proposed mechanism is geometrically simple: smooth elastic turbulence creates steep polymer fronts, forming in near-plane-strain geometry; their normals align with the compressive direction, and their amplitude grows while their thickness shrinks.  
The theory resolves the long-standing spectral-selection problem left open by Fouxon and Lebedev \cite{fouxon_lebedev_2003k}, providing a physical mechanism that uniquely determines both exponents. 

The mechanism proposed here may extend well beyond the homogeneous, triply periodic flow considered in this work. In recent years, considerable attention has been devoted to identifying and characterizing the coherent structures that emerge in elastic turbulent planar channel flows, including the so-called narwhal or center-mode structures \cite{page_dubief_kerswell_2020a, buza_beneitez_page_kerswell_2022a, morozov_2022a, lellep_linkmann_morozov_2023a, foggi-rota_amor_le-clainche_rosti_2024a, lewy_kerswell_2026a, zhu_kerswell_2026a, amor_corrochano_soligo_le-clainche_rosti_2026a, goffin_dubief_terrapon_2026a}. These flow structures exhibit precisely the kind of near-plane-strain hyperbolic topology that underlies the mechanism proposed here. Comparable hyperbolic structures have also been identified in porous-media flows, free jets, and other elastic turbulent configurations, despite their very different forcing mechanisms and boundary conditions, suggesting that the underlying dynamics may be more universal than previously anticipated. 

More broadly, these results suggest that elastic frontogenesis may represent the analogue of vortex stretching in inertial turbulence. Whereas vortex stretching transfers energy across scales and underpins the universal spectral structure of inertial turbulence, elastic frontogenesis continually sharpens polymer gradients and gives rise to the spectra of elastic turbulence. If confirmed across diverse viscoelastic flows, elastic frontogenesis would provide the missing organizing principle connecting coherent structures, nonlinear dynamics and spectral scaling in elastic turbulence.

\acknowledgments{\section*{Acknowledgments}
M.E.R. was supported by the Okinawa Institute of Science and Technology Graduate University (OIST) with subsidy funding from the Cabinet Office, Government of Japan. M.E.R. also acknowledges funding from the Japan Society for the Promotion of Science (JSPS), grant 24K17210 and 24K00810. The authors acknowledge the computer time provided by the Scientific Computing \& Data Analysis section of the Core Facilities at OIST, and by HPCI, under the Research Project grants hp250035 and hp260009.
}


%

\newpage

\section*{End Matter}

\subsection{Direct numerical simulations} \label{sec:em-methods}
The governing equations are solved in a triply periodic cubic domain of size $L=2\pi$ using a second-order central finite-difference scheme on a uniform Cartesian grid with $N$ collocation points in each direction. Time integration is performed using a second-order Adams--Bashforth scheme with a time step satisfying $\Delta t \approx 10^{-5}\tau_p$. We use the log-conformation approach to handle the stress constitutive equation, which has been shown to be reliable at large $De$, and hence we do not add any artificial stress diffusion, and avoid any spurious artifacts the latter can introduce \cite{fattal_kupferman_2004a, yerasi_picardo_gupta_vincenzi_2024a}.

The turbulent flow is sustained using the ABC forcing with $\boldsymbol{F_{ext}} = (\sin z + \cos y) \hat{\boldsymbol{x}} + (\sin x + \cos z) \hat{\boldsymbol{y}} + (\sin y + \cos x) \hat{\boldsymbol{z}}$ \cite{arnold_2013a}, allowing us to reach a statistically steady state over which statistics are accumulated. The forcing has amplitude $F$ and is active only at the first wavenumber $k_f = k_\textrm{min}$. The Reynolds number $Re = \rho_f U_F / \left( k_F \mu_f \right)$ is equal to approximately $11$ (where $U_F$ the forcing velocity scale $\sqrt{F/k_F}$), the Deborah number $De=\tau_p U_F k_F$ is fixed to a value $6$, safely larger than unity, and the viscosity ratio $\mu_f/\left( \mu_f +\mu_p \right)$ to $0.9$ for dilute polymer solutions. For the FENE-P model, we choose the maximum polymer extensibility $L_m$ to be equal to $50$. 

The Eulerian statistics presented in this work are computed from $10$ statistically independent flow snapshots, separated by at least $\tau_p$. The Lagrangian statistics are obtained by seeding the flow with $10^5$ passive tracers, and storing their time histories at every time-step over a period of several relaxation times. All simulations are performed using the in-house solver Fujin \cite{rosti_2026a}.

\subsection{From polymer to fluid spectra} \label{sec:em-polymer1}
The kinetic exponent follows from the small-scale force balance. Assume that in the scaling range inertial terms and external forcing are negligible, an assumption verified by the spectral energy budget reported in the Supplementary Material \cite{supplementary}, so that
\begin{equation}
0 \simeq
-\boldsymbol{\nabla}p
+\mu_f\nabla^2\boldsymbol{u}
+\frac{\mu_p}{\tau_p}
\boldsymbol{\nabla}\cdot
\left(f_1\boldsymbol{C}-f_2\boldsymbol{I}\right).
\end{equation}
After Fourier transforming and projecting onto incompressible modes, we obtain, for $\boldsymbol{k}\neq\boldsymbol{0}$,
\begin{equation}
\mu_f k^2\hat{u}_i(\boldsymbol{k})
\simeq
\frac{\mu_p}{\tau_p}\,
iP_{i\ell}(\boldsymbol{k})k_j
\widehat{f_1C_{\ell j}}(\boldsymbol{k}),
\qquad
k=|\boldsymbol{k}|,
\end{equation}
where $ P_{i\ell}(\boldsymbol{k}) = \delta_{i\ell}-k_i k_\ell / k^2$ is the solenoidal projector. The isotropic term proportional to $f_2$ does not contribute at non-zero wavenumber, since $f_2$ is constant. We define the solenoidally active polymer-stress component $ \hat{\Theta}_i(\boldsymbol{k}) \equiv iP_{i\ell}(\boldsymbol{k}) k_j \widehat{f_1C_{\ell j}}(\boldsymbol{k})/k$, and its shell-integrated spectrum
\begin{equation}
E_\Theta(k)
\equiv
\int_{|\boldsymbol{q}|=k}
\left\langle
\hat{\Theta}_i(\boldsymbol{q},t)
\hat{\Theta}_i^*(\boldsymbol{q},t)
\right\rangle
\,dS_q.
\end{equation}
The projected force balance can then be written as
$ \mu_f k^2\hat{u}_i(\boldsymbol{k}) \simeq \frac{\mu_p}{\tau_p}\, k\hat{\Theta}_i(\boldsymbol{k})$, and therefore
\begin{equation}
E_f(k)
\simeq
\left(\frac{\mu_p}{\mu_f\tau_p}\right)^2
\frac{E_\Theta(k)}{k^2}.
\label{eq:Ef_Etheta}
\end{equation}

\begin{figure*}[t]
    \centering
    \includegraphics[width=0.70\textwidth]{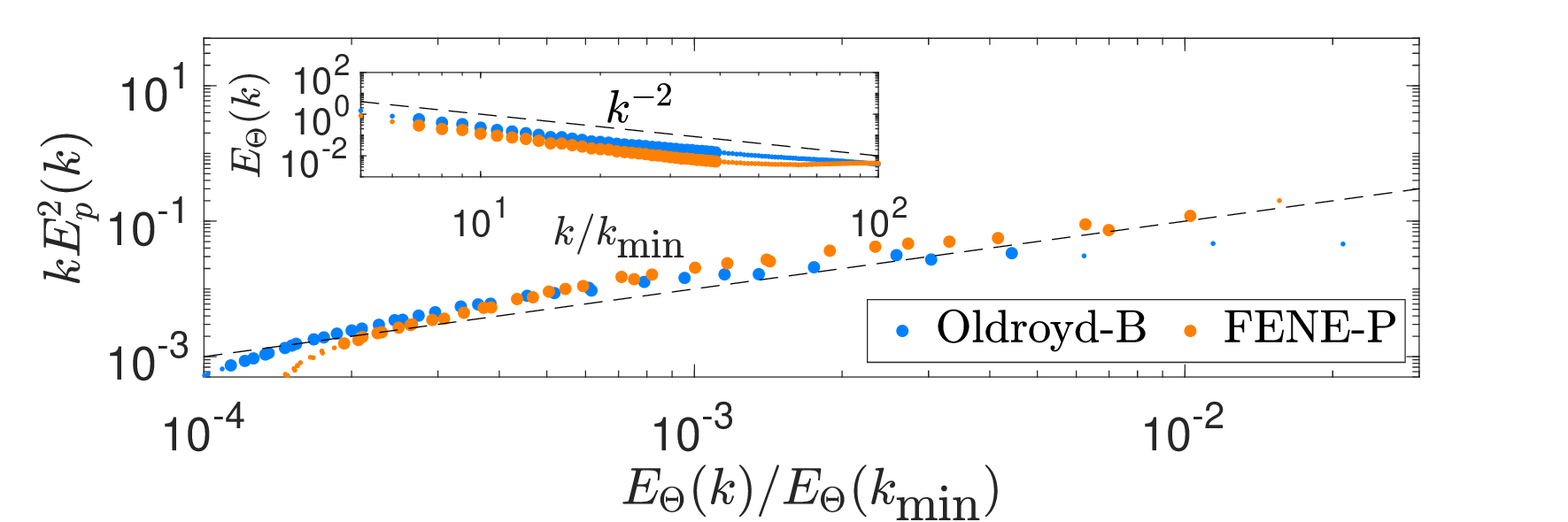}
    \caption{\textbf{Direct verification of the scale-local closure.} $kE_p^2(k)$ (normalized by $k_\textrm{min}E_p^2(k_\textrm{min})$) plotted against normalized $E_\Theta(k)$ for the Oldroyd-B (blue) and FENE-P (orange) models at $N=512$. The dashed line denotes a line with unitary slope. The collapse over the displayed fitted range (large symbols) is consistent with $E_\Theta(k)\propto kE_p^2(k)$. The inset shows $E_\Theta(k)$ versus $k/k_\textrm{min}$, with a $k^{-2}$ reference slope. For FENE-P, $E_\Theta$ is computed from the complete product $f_1\boldsymbol{C}$.}
    \label{fig:peterlin}
\end{figure*}

It remains to relate $E_\Theta(k)$ to the polymer spectrum. The polymer stress is quadratic in $\boldsymbol{B}$, so that each of its Fourier modes is a convolution of two modes of $\boldsymbol{B}$ and, for a spectrum as steep as the present one, non-local products of large-scale and small-scale modes could in principle dominate. For sheet-like fronts, however, these leading non-local contributions are geometrically inactive: the amplified conformation is tangent to the sheet, tangential stresses exert no traction along the front normal, and a purely normal traction is removed by the solenoidal projector. The solenoidally active stress at wavenumber $k$ is therefore generated locally in scale, by modes of $\boldsymbol{B}$ of comparable wavenumber, so that
\begin{equation}
E_\Theta(k)\sim kE_p^2(k),
\label{eq:projected_relation}
\end{equation}
up to an order-one geometric coefficient $\gamma$ independent of $k$ (see Supplementary Material for the derivation, including the FENE-P case, in which the polymer stress carries the additional scalar factor $f_1$).

Combining equations~\eqref{eq:Ef_Etheta} and~\eqref{eq:projected_relation} gives
\begin{equation}
E_f(k)\sim\frac{E_p^2(k)}{k}.
\label{eq:Ef_Ep}
\end{equation}
Thus, if $E_p(k)\sim k^{-\chi}$ and $E_f(k)\sim k^{-\xi}$, the scale-local front closure predicts $ \xi=2\chi+1$. Using the frontogenesis result $\chi=2-\bar a$, this yields $ \boxed{\xi=5-2\bar a}$. In particular, $\bar a=1/2$ gives the joint prediction
\begin{equation}
E_p(k)\sim k^{-3/2},
\qquad
E_\Theta(k)\sim k^{-2},
\qquad
E_f(k)\sim k^{-4}.
\end{equation}

Figure~\ref{fig:peterlin} tests the intermediate prediction~\eqref{eq:projected_relation} directly, rather than using it as an input. Here, $E_\Theta(k)$ is computed from the complete polymer stress entering the momentum equation, including its full Fourier convolution, tensorial contraction, solenoidal projection and, for FENE-P, the spatially varying factor $f_1$. Over the fitted scaling range, the DNS data collapse onto the unit-slope line, verifying Eq.~\eqref{eq:projected_relation} for both the Oldroyd-B and FENE-P models and showing that the effective coefficient $\gamma$ is scale-independent within the observed range. The same direct measurement gives $E_\Theta(k)\sim k^{-2}$ (inset of Fig.~\ref{fig:peterlin}), consistently with the joint theoretical prediction above. Note that, while our  numerical simulations of ET rely on Eulerian models employing constitutive equations, Lagrangian approaches have also been used in the past \cite{watanabe_gotoh_2014a, serafini_2025a}, that have shown the same slope for $E_f(k)$ and $E_\Theta(k)$ \cite{serafini_2025a}, proving once more the robustness of the results discussed here.

\end{document}


\preprint{APS/123-QED}

\title{Supplementary Material \\ Elastic frontogenesis sets the spectra of smooth elastic turbulence}

\author{Marco Edoardo Rosti\orcidA{}}
\email{marco.rosti@oist.jp}
\affiliation{Complex Fluids and Flows Unit, Okinawa Institute of Science and Technology Graduate University, Okinawa 904-0495, Japan}
\author{Andrea Mazzino\orcidB{}}
\email{andrea.mazzino@unige.it}
\affiliation{Department of Civil, Chemical and Environmental Engineering, University of Genova, Via Montallegro 1, Genova, 16145, Genova, Italy \\
            Istituto Nazionale di Fisica Nucleare, Sezione di Genova, Via Dodecaneso 33, Genova, 16146, Italy}

\maketitle

\setcounter{table}{0}
\makeatletter 
\renewcommand{\thetable}{S\@arabic\c@table}
\makeatother

\setcounter{figure}{0}
\makeatletter 
\renewcommand{\thefigure}{S\@arabic\c@figure}
\makeatother

\setcounter{equation}{0}
\makeatletter 
\renewcommand{\theequation}{S\@arabic\c@equation}
\makeatother

\section{Energy balance}

The relevant energy flux equation has already been derived several times in the past; here we follow the notation and convention of \citet{abdelgawad_cannon_rosti_2023a}. We start from the Fourier transform of the momentum equation
\begin{equation}
	\rho_f \partial_t \hat{\boldsymbol{u}}
	+\hat{\boldsymbol{G}}
	=
	-i\boldsymbol{k}\hat{p}
	-\mu_\text{f}k^{2}\hat{\boldsymbol{u}}
	+\hat{\boldsymbol{F}}_\text{pol}
	+\hat{\boldsymbol{F}}_\text{ext},
\end{equation}
where $\hat{\boldsymbol{G}}$ is the Fourier coefficient of the nonlinear convective term, $i$ is the imaginary unit, and
\begin{equation}
\widehat{F}_{{\rm pol},r}(\boldsymbol{k})
=
i\frac{\mu_p}{\tau_p}
k_j
\widehat{
\left(f_1C_{rj}-f_2\delta_{rj}\right)
}(\boldsymbol{k}).
\end{equation}
We define the modal kinetic energy as
\begin{equation}
\hat{E}(\boldsymbol{k},t)
\equiv
\frac{1}{2}
\hat{\boldsymbol{u}}(\boldsymbol{k},t)
\cdot
\hat{\boldsymbol{u}}^{*}(\boldsymbol{k},t).
\end{equation}
A similar momentum equation holds for the complex conjugate $\hat{\boldsymbol{u}}^{*}$. When the equation for $\hat{\boldsymbol{u}}$ is multiplied by $\hat{\boldsymbol{u}}^{*}$, the pressure term
$-i\boldsymbol{k}\cdot\hat{\boldsymbol{u}}^{*}\hat{p}$
vanishes because of the incompressibility constraint, while the viscous term
$-\mu_\text{f}k^{2}\hat{\boldsymbol{u}}\cdot\hat{\boldsymbol{u}}^{*}$
can be written as $-2\mu_\text{f}k^{2}\hat{E}$. The same holds when the conjugate momentum equation is multiplied by $\hat{\boldsymbol{u}}$. Adding the two resulting equations and dividing by $2$ gives the evolution equation for the modal kinetic energy,
\begin{equation}
\label{eq:Espec}
	\rho_f\partial_t\hat{E}(\boldsymbol{k},t)
	=
	\hat{N}(\boldsymbol{k},t)
	+\hat{D}(\boldsymbol{k},t)
	+\hat{P}(\boldsymbol{k},t)
	+\hat{S}(\boldsymbol{k},t),
\end{equation}
where
\begin{figure}[tb]
	\includegraphics[width=0.99\textwidth]{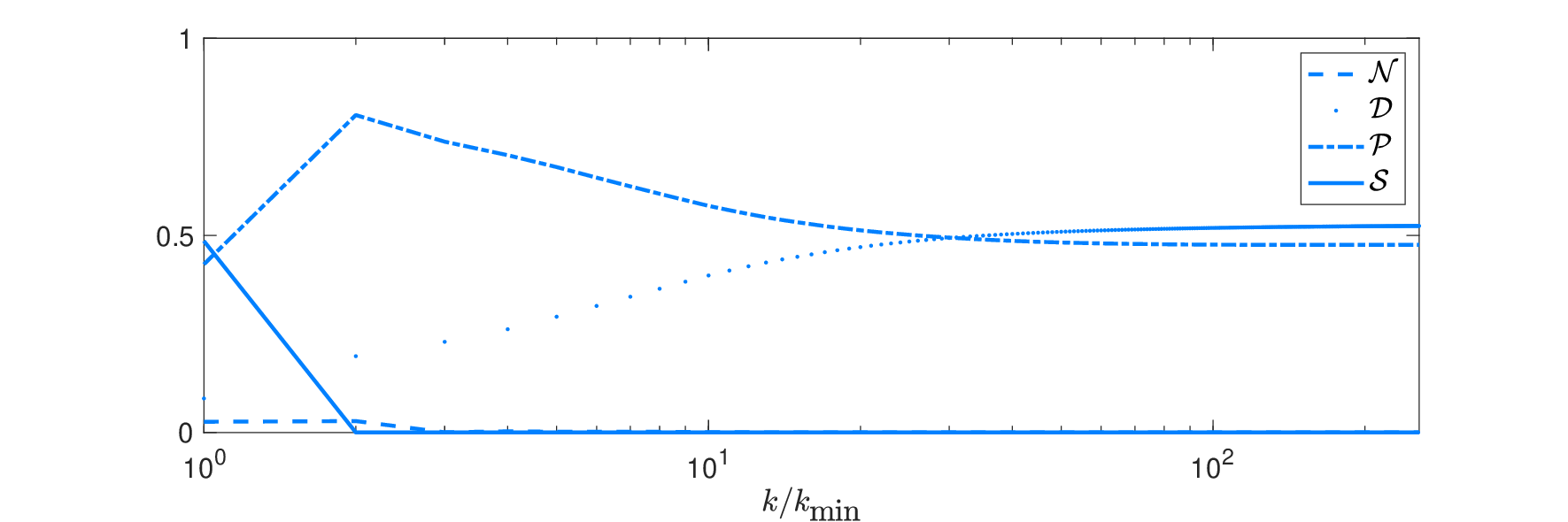}
    \includegraphics[width=0.99\textwidth]{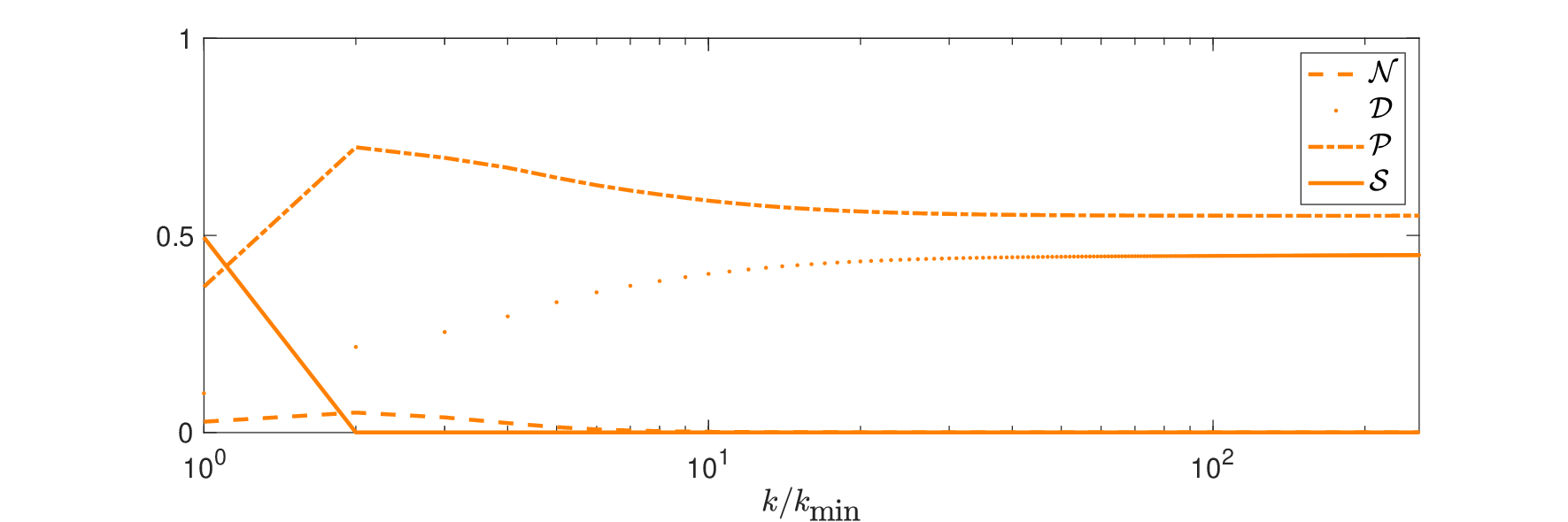}
	\caption{\textbf{Energy balance.} The inertial ($\mathcal{N}$), viscous dissipation ($\mathcal{D}$), polymeric ($\mathcal{P}$), and driving force ($\mathcal{S}$) contributions to the energy balance for the (top) Oldroyd-B and (bottom) FENE-P models with $N=512$. All terms are normalised by $\varepsilon_f+\varepsilon_p$.}
	\label{fig:figS1}
\end{figure}
\begin{equation}
\hat{N}
=
-\frac{1}{2}
\left(
\hat{\boldsymbol{G}}\cdot\hat{\boldsymbol{u}}^{*}
+
\hat{\boldsymbol{G}}^{*}\cdot\hat{\boldsymbol{u}}
\right)
\end{equation}
is the contribution from the nonlinear convective term,
\begin{equation}
\hat{D}
=
-2\mu_\text{f}k^{2}\hat{E}
\end{equation}
is the viscous contribution,
\begin{equation}
\hat{P}
=
\frac{1}{2}
\left(
\hat{\boldsymbol{F}}_\text{pol}\cdot\hat{\boldsymbol{u}}^{*}
+
\hat{\boldsymbol{F}}_\text{pol}^{*}\cdot\hat{\boldsymbol{u}}
\right)
\end{equation}
is the polymeric contribution, and
\begin{equation}
\hat{S}
=
\frac{1}{2}
\left(
\hat{\boldsymbol{F}}_\text{ext}\cdot\hat{\boldsymbol{u}}^{*}
+
\hat{\boldsymbol{F}}_\text{ext}^{*}\cdot\hat{\boldsymbol{u}}
\right)
\end{equation}
is the contribution from the external forcing. After integrating each term over the spherical shell $|\boldsymbol{k}|=k$, we obtain the equation for the one-dimensional energy spectrum $E(k,t)$,
\begin{equation}
\label{eq:Espec_avg}
	\rho_f\partial_tE(k,t)
	=
	N(k,t)+D(k,t)+P(k,t)+S(k,t).
\end{equation}
For a statistically stationary flow,
$\partial_t\langle E(k,t)\rangle=0$.
In the following, we omit the statistical averaging brackets. Integrating from $k$ to infinity, we obtain the energy-transfer balance
\begin{equation}
\label{eq:E_balance}
	0
	=
	\mathcal{N}(k)
	+\mathcal{D}'(k)
	+\mathcal{P}'(k)
	+\mathcal{S}(k),
\end{equation}
where
\begin{equation}
\mathcal{N}(k)
\equiv
\int_k^\infty N(q)\,\mathrm{d}q,
\qquad
\mathcal{D}'(k)
\equiv
\int_k^\infty D(q)\,\mathrm{d}q,
\end{equation}
\begin{equation}
\mathcal{P}'(k)
\equiv
\int_k^\infty P(q)\,\mathrm{d}q,
\qquad
\mathcal{S}(k)
\equiv
\int_k^\infty S(q)\,\mathrm{d}q
\end{equation}
represent the contributions to the spectral power balance from the nonlinear convective term, fluid viscosity, polymer stress, and external forcing, respectively. The cumulative viscous contribution can be written as
\begin{equation}
\mathcal{D}(k)
\equiv
-\int_0^kD(q)\,\mathrm{d}q
=
\mathcal{D}'(k)+\varepsilon_f,
\end{equation}
where
\begin{equation}
\varepsilon_f
\equiv
-\int_0^\infty D(q)\,\mathrm{d}q
\end{equation}
is the rate of energy dissipated by fluid viscosity. Similarly, the cumulative polymeric contribution can be written as
\begin{equation}
\mathcal{P}(k)
\equiv
-\int_0^kP(q)\,\mathrm{d}q
=
\mathcal{P}'(k)+\varepsilon_p,
\end{equation}
where
\begin{equation}
\varepsilon_p
\equiv
-\int_0^\infty P(q)\,\mathrm{d}q
\end{equation}
is the net rate of energy transferred from the fluid to the polymers. Substituting these relations into equation~\eqref{eq:E_balance}, we obtain the final balance equation,
\begin{equation}
\varepsilon_f+\varepsilon_p
=
\mathcal{N}(k)
+\mathcal{D}(k)
+\mathcal{P}(k)
+\mathcal{S}(k).
\end{equation}

Figure~\ref{fig:figS1} shows the energy balance for the two models considered with $N=512$. As is clear from the plot, the nonlinear energy transfer $\mathcal{N}$ is negligible, showing that nonlinear inertial transfer plays a negligible role in the spectral energy balance. Note, however, that $\mathcal{N}$ is larger for the FENE-P model, consistently with shear-thinning enhancing nonlinear inertial effects \cite{amor_soligo_mazzino_rosti_2024a,rosti_2025a}. The forcing contribution $\mathcal{S}$ is nonzero only at the forcing wavenumber $k=1$. At all other scales, ET is characterised by a balance between the viscous contribution $\mathcal{D}$ and the polymeric contribution $\mathcal{P}$ \cite{singh_perlekar_mitra_rosti_2024a}.

\section{Choice of the condition set and threshold}

\begin{figure}[h]
	\includegraphics[width=0.32\textwidth]{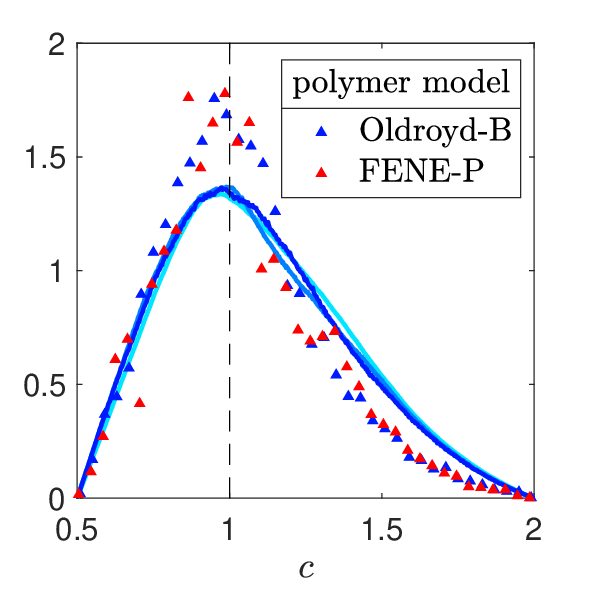}
	\includegraphics[width=0.32\textwidth]{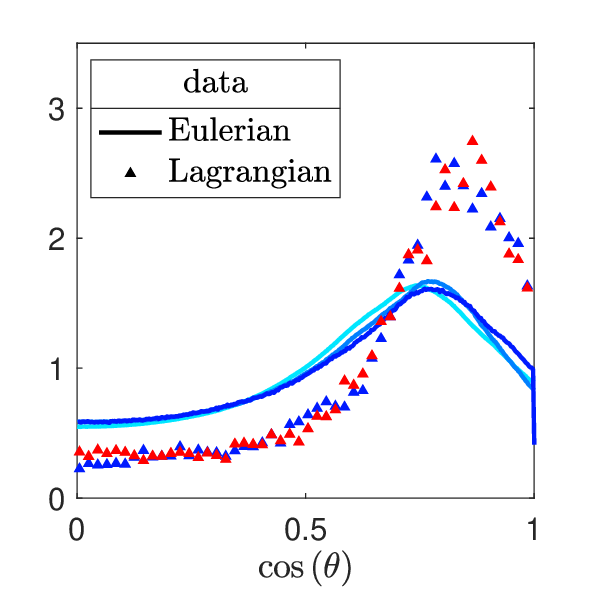}	
	\includegraphics[width=0.32\textwidth]{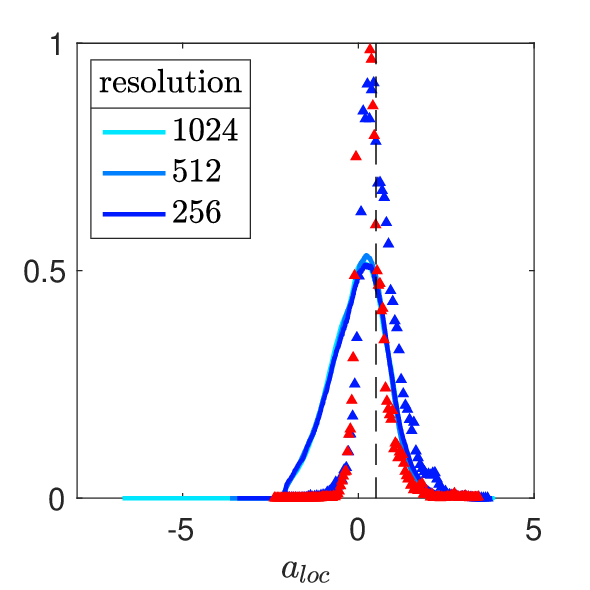}
	\caption{\textbf{Effect of the condition set on the pdf of the local diagnostics.} (left) $c=-s_3/s_1$, (middle) $\cos \left( \theta \right) = |\hat{\bm n}\cdot\bm e_3|$, and (right) $a_{\rm loc}$. The blue and red colors are used to distinguish the results from the Oldroyd-B and FENE-P models, respectively. The lines are unconditioned Eulerian data, with color tone indicating different grid resolutions $N$. The symbols indicate Lagrangian data conditioned on $\mathcal S_f$, with threshold $\Theta_f$ chosen as the quantile 0.99 ($\bigtriangleup$).}
	\label{fig:figS3}
\end{figure}

\begin{figure}[h]
	\includegraphics[width=0.32\textwidth]{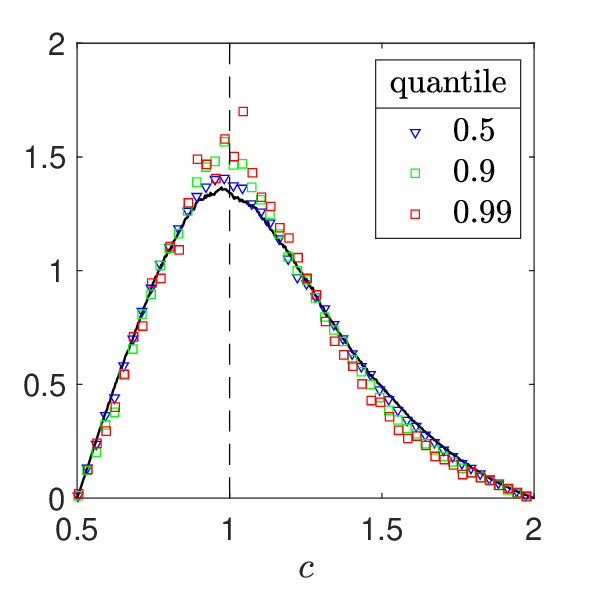}
	\includegraphics[width=0.32\textwidth]{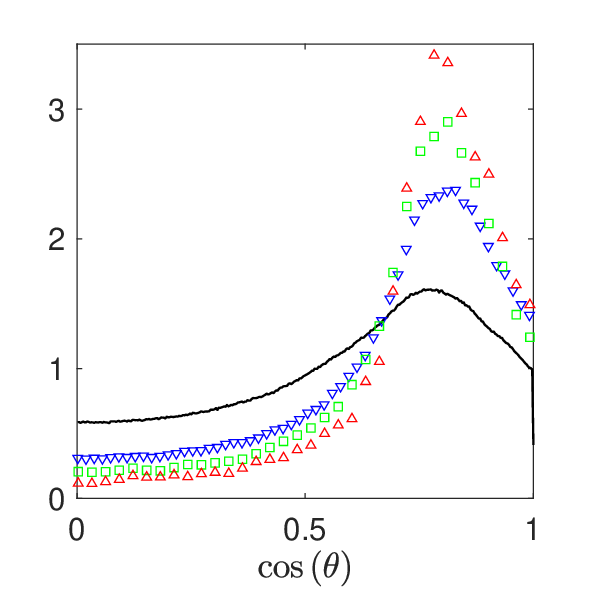}	
	\includegraphics[width=0.32\textwidth]{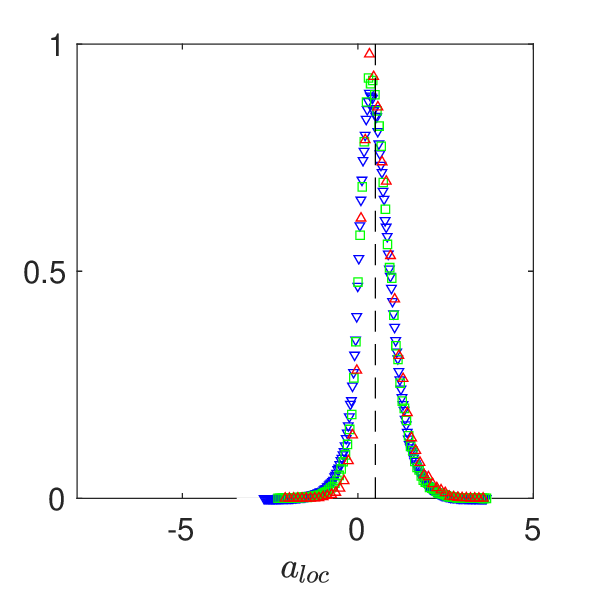}
	\caption{\textbf{Effect of the threshold on the pdf of the local diagnostics.} (left) $c=-s_3/s_1$, (middle) $\cos \left( \theta \right) = |\hat{\bm n}\cdot\bm e_3|$, and (right) $a_{\rm loc}$ for the Oldroyd-B model. The black line represents the unconditioned Eulerian data, while the symbols indicate Lagrangian data conditioned on $\mathcal S_p$, with thresholds $\Theta_p$ chosen as the quantile 0.5 ($\bigtriangledown$), 0.9 ($\square$), and 0.99 ($\bigtriangleup$).}
	\label{fig:figS2}
\end{figure}

We now assess the effect of the condition set used to assess the local diagnostics reported in the main manuscript. In particular, figure~\ref{fig:figS3} reproduces the one in the manuscript, with the conditioning done on $\mathcal{S}_f$.  Even in this case, the same conclusions drawn in the main manuscript still hold, with $c \approx 1$, $\cos \left( \theta \right) \gg 1/2$, and $a_{\rm loc} \approx 1/2$.

Next, we assess in figure~\ref{fig:figS2} the sensitivity of the results to the choice of the threshold $\Theta_p$. The figure reports the results for three different values, corresponding to the the quantiles 0.5, 0.9, and 0.99. The figure clearly shows that, when the threshold is low, the distributions converge towards the unconditioned Eulerian distribution, while the peaks of all the curves sharpen as the threshold is increased.

\begin{figure}[h]
    \includegraphics[width=0.30\textwidth]{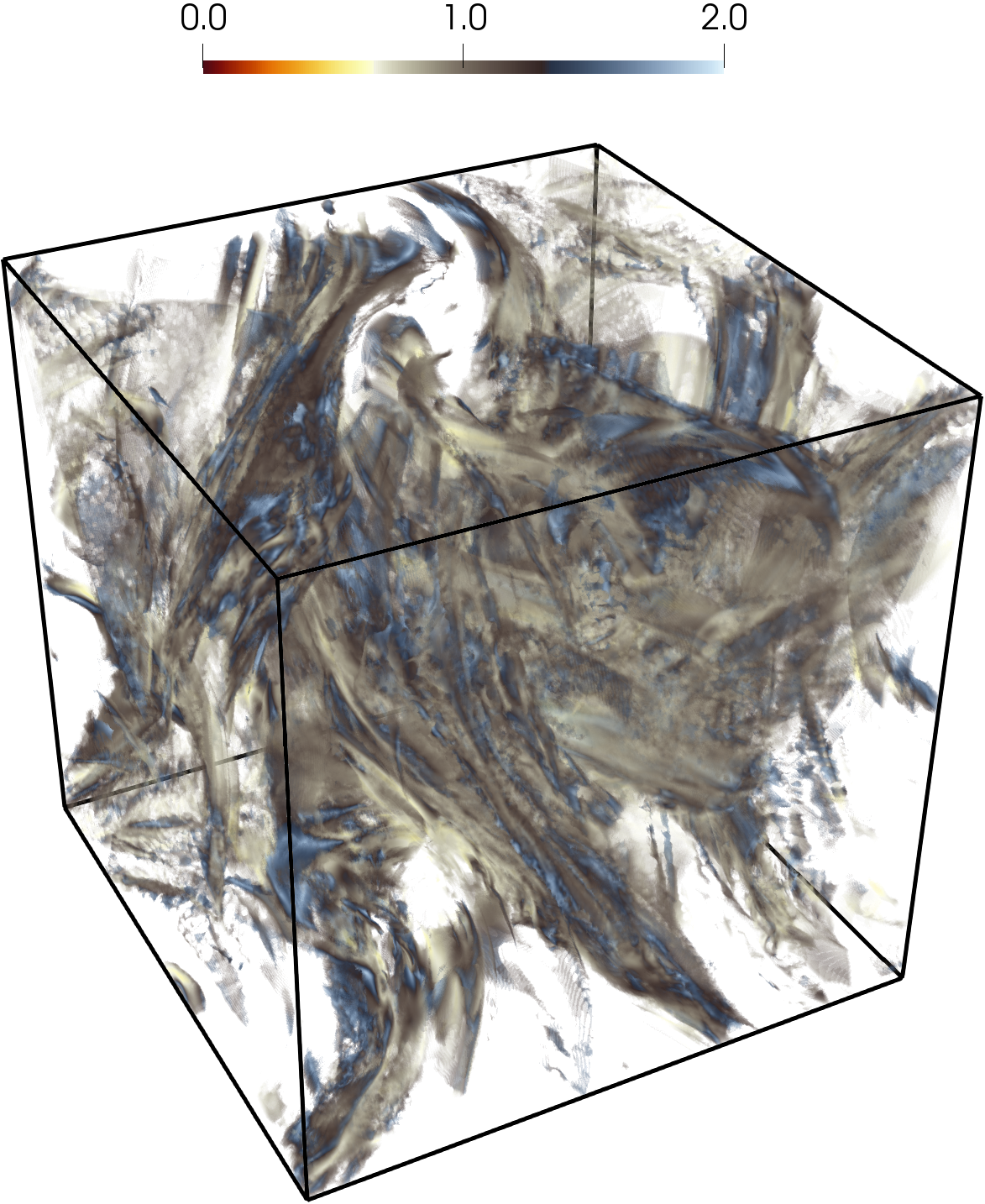}
	\includegraphics[width=0.30\textwidth]{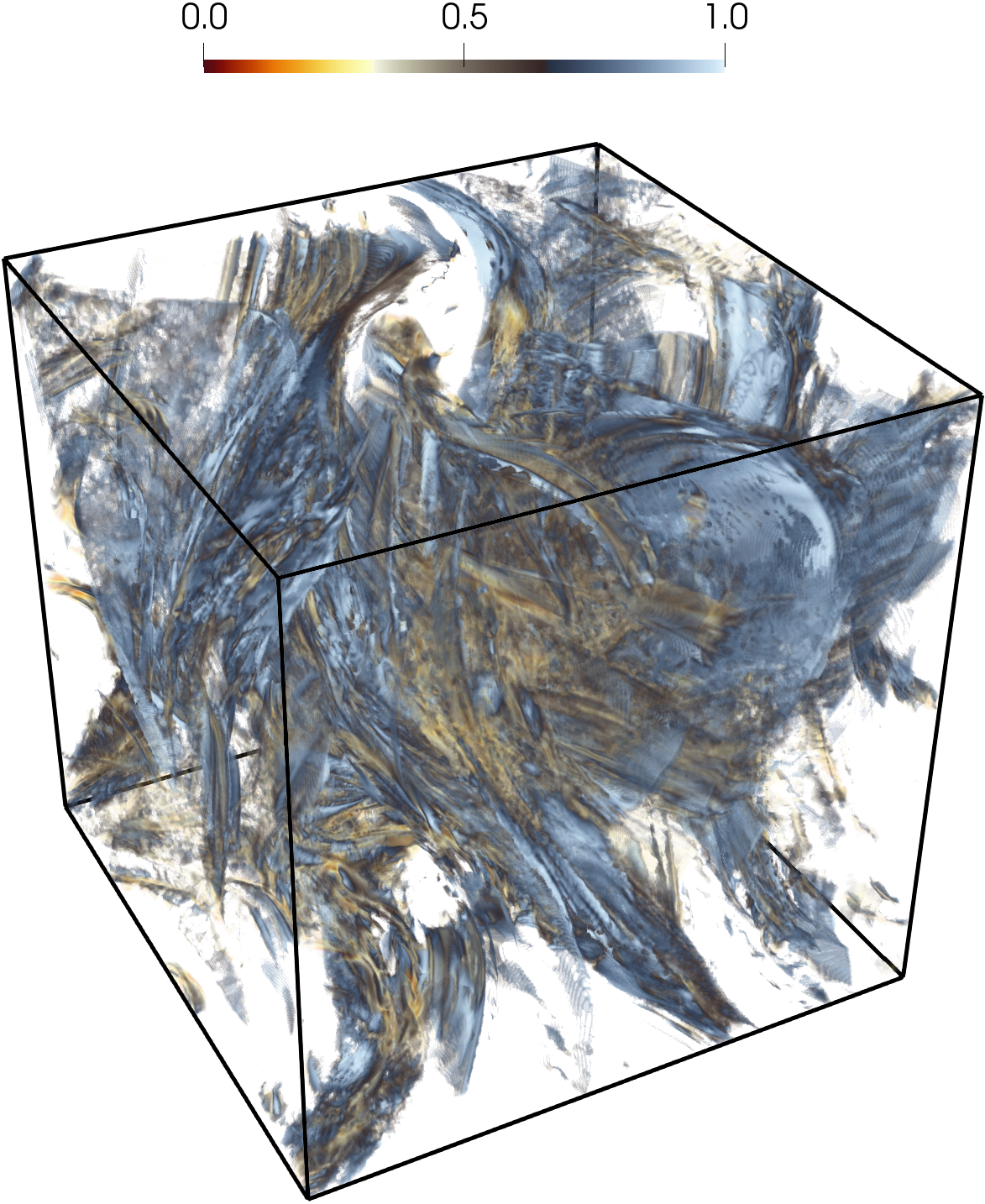}
	\includegraphics[width=0.30\textwidth]{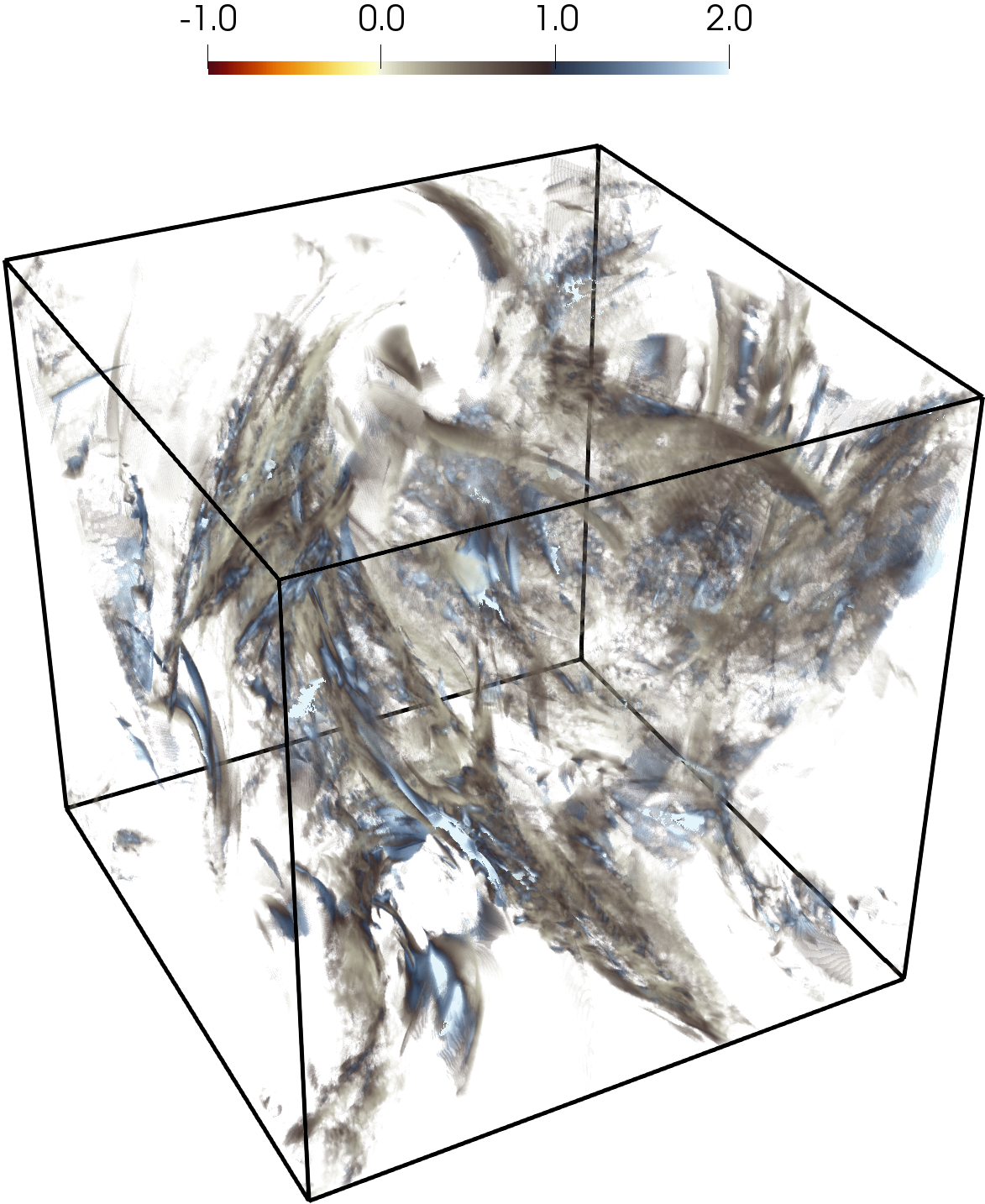}
	\caption{\textbf{Polymer-stress sheets in real space.} Instantaneous visualization of the local diagnostics on the shown colorbar (the whole range of colors is not used in all the panels). (left) $c=-s_3/s_1$, with expected value in gray ($1$). (middle) $\cos \left( \theta \right) = |\hat{\bm n}\cdot\bm e_3|$, with expected value in blue ($1$). (right) $a_{\rm loc}$, with expected value in gray ($0.5$). The represented volumes are conditioned on $\mathcal S_p$, with thresholds $\Theta_p$ chosen as the quantile 0.99. The right panel is restricted to growing regions, $g>0$.}
	\label{fig:fig5}
\end{figure}

Finally, figure~\ref{fig:fig5} shows an instantaneous visualisation of the local diagnostics conditioned on $\mathcal S_p$, visually confirming the presence of polymer sheets.

\section{Scale locality of the solenoidally active polymer stress} \label{sec:em-polymer2}
Here we derive the scale-local closure used in the main manuscript, Eq.~18, relating the spectrum of the solenoidally active polymer stress to the polymer conformation spectrum. Let $\boldsymbol{B}^{[k]}$ and $\boldsymbol{\Theta}^{[k]}$ denote the contributions to the corresponding fields from a logarithmic band of wavenumbers of order $k$, and let $\boldsymbol{B}^{<k}$ denote the low-pass component of $\boldsymbol{B}$ containing wavenumbers well below $k$. We define the characteristic squared amplitudes of the band-filtered fields as 
\begin{equation}
\begin{split}
b_k^2
\equiv
\left\langle
B_{ij}^{[k]}B_{ij}^{[k]}
\right\rangle
\sim kE_p(k),\\
\theta_k^2
\equiv
\left\langle
\Theta_i^{[k]}\Theta_i^{[k]}
\right\rangle
\sim kE_\Theta(k).
\label{eq:band_amplitudes}
\end{split}
\end{equation}
For Oldroyd-B, $\boldsymbol{C}=\boldsymbol{B}\boldsymbol{B}^{T}$, so that each Fourier mode of the polymer stress is generated by a convolution of two Fourier modes of $\boldsymbol{B}$. For a generic quadratic product, however, the non-local low--high terms $ \boldsymbol{B}^{<k} \big(\boldsymbol{B}^{[k]}\big)^T + \boldsymbol{B}^{[k]} \big(\boldsymbol{B}^{<k}\big)^T $ could dominate over interactions between modes of comparable wavenumber, particularly for a steep spectrum.

In the present case, these leading non-local contributions are suppressed by the combination of the front geometry and the solenoidal projection. For a locally planar front, the wavevector associated with its rapid variation is approximately normal to the sheet, $\hat{\boldsymbol{k}} \simeq \hat{\boldsymbol{n}} \simeq \boldsymbol{e}_3$, whereas the amplified conformation is predominantly tangent to the sheet. In the ideal planar and locally coaxial limit, the leading low--high contribution therefore satisfies 
\begin{equation}
\boldsymbol{P}(\boldsymbol{k})
\left[
\boldsymbol{B}^{<k}
\big(\boldsymbol{B}^{[k]}\big)^T
+
\boldsymbol{B}^{[k]}
\big(\boldsymbol{B}^{<k}\big)^T
\right]
\hat{\boldsymbol{k}}
\simeq
\boldsymbol{0}.
\label{eq:nonlocal_cancellation}
\end{equation}
Indeed, tangential stress components exert no normal traction, whereas a purely normal traction is parallel to $\hat{\boldsymbol{k}}$ and is removed by the solenoidal projector. Thus, although the low--high terms could dominate for a generic quadratic product, their leading contribution is inactive in the projected momentum equation.

Equation~\eqref{eq:nonlocal_cancellation} describes the ideal planar and locally coaxial limit. We further assume that departures from this geometry do not restore a low--high contribution that dominates, in scaling order, over interactions between modes of order $k$. Under this closure, the leading solenoidally active contribution is scale local, and we write
\begin{equation}
\theta_k
\sim
\gamma b_k^2,
\label{eq:local_closure}
\end{equation}
where $\gamma$ is a dimensionless, order-one coefficient accounting for the front geometry, the tensorial contraction, and the solenoidal projection. Statistical self-similarity here means simply that, once their thickness and amplitude are rescaled out, fronts have the same typical shape, orientation, and tensorial structure throughout the scaling range. The efficiency with which the front-scale stress produces a solenoidal force is therefore independent of $k$, so that $\gamma$ introduces no additional spectral exponent. Combining equations~\eqref{eq:band_amplitudes} and~\eqref{eq:local_closure} gives $kE_\Theta(k) \sim \gamma^2 \left[kE_p(k)\right]^2$, and hence Eq.~18 of the main text. Equivalently, the scale-local front closure predicts $ E_\Theta(k) / \left( kE_p^2(k) \right) \sim  \gamma^2 = \mathrm{const.}$ over the front-dominated scaling range.

For FENE-P, $E_\Theta(k)$ is evaluated from the Fourier transform of the complete product $ f_1\boldsymbol{C} = f_1\boldsymbol{B}\boldsymbol{B}^{T}$. Since $f_1$ is a scalar, it changes the local stress amplitude but not its tensorial directions, so that the same geometrical suppression of the leading projected low--high contribution applies. Its spatial fluctuations may change the numerical value of the effective coefficient $\gamma$. The corresponding prediction is that, over the intermediate front-dominated range, they do not make $\gamma$ scale-dependent and therefore do not introduce an additional spectral exponent. This does not require $f_1$ to be spatially constant or to be replaced by a scale-independent approximation.

%